\documentstyle[12pt,epsfig]{article}                                                      
                                                      
\newlength{\dinwidth}                                                      
\newlength{\dinmargin}                                                      
\def\lapproxeq{\lower .7ex\hbox{$\;\stackrel{\textstyle                                                      
<}{\sim}\;$}}                                                      
\def\gapproxeq{\lower .7ex\hbox{$\;\stackrel{\textstyle                                                      
>}{\sim}\;$}}                                                      
\def\be{\begin{equation}}                                                      
\def\ee{\end{equation}}                                                      
\def\bea{\begin{eqnarray}}                                                      
\def\eea{\end{eqnarray}}

\newcounter{bean}                    
                                                      
\begin{document}                                                      
%\titlepage                                                      
\begin{flushright}                                                      
DTP/00/34 \\                                                      
May 2000 \\                                                      
\end{flushright}                                                      
                                                      
\vspace*{2cm}                                                      
                                                      
\begin{center}                                                      
{\Large \bf The unintegrated gluon distribution \\
from the CCFM equation} \\             
             
%\vspace*{0.5cm}             
%{\Large \bf ???}                                                      
                                                      
\vspace*{1cm}                                                      
M.A. Kimber$^a$, J. Kwiecinski$^{a,b}$, A.D. Martin$^a$ and A.M. Stasto$^{a,b}$  \\                                                      
                                                     
\vspace*{0.5cm}                
$^a$ Department of Physics, University of Durham, Durham, DH1 3LE \\                                                     
$^b$ H.~Niewodniczanski Institute of Nuclear Physics, ul.~Radzikowskiego 152,                                   
Krakow, Poland \\         
\end{center}                                                      
                                                      
\vspace*{1cm}                                                      
                                                      
\begin{abstract}                                                      
The gluon distribution $f (x, k_t^2, \mu^2)$, unintegrated over the transverse momentum        
$k_t$ of the gluon, satisfies the angular-ordered CCFM equation which interlocks        
the dependence on the scale $k_t$ with the scale $\mu$ of the probe.  We show how, to        
leading $\log$ accuracy, the equation can be simplified to a single scale problem.  In        
particular we demonstrate how to determine the two-scale unintegrated distribution $f (x,        
k_t^2, \mu^2)$ from knowledge of the integrated gluon obtained from a unified scheme    
embodying both BFKL $(\log (1/x))$ and DGLAP $(\log \mu^2)$ evolution.        
\end{abstract}                                            
             
\section{Introduction}           
Deep inelastic electron-proton scattering is described in terms of scale dependent parton        
distributions $q (x, \mu^2)$ and $g (x, \mu^2)$.  For less inclusive processes it is however        
necessary to consider distributions unintegrated over the transverse momentum $k_t$ of the        
parton, which for the gluon, for example, we denote by $f (x, k_t^2, \mu^2)$.  These        
distributions depend on two hard scales; $k_t$ and the hard scale $\mu$ of the probe.  The        
(conventional) integrated gluon distribution is given by         
\be         
\label{eq:a1}         
xg (x, \mu^2) \; = \; \int^{\mu^2} \frac{dk_t^2}{k_t^2} \: f (x, k_t^2, \mu^2).         
\ee         
Unintegrated distributions are required to describe measurements where transverse momenta        
are exposed explicitly.  For example to describe the $p_T$ spectrum of prompt photons        
produced in high energy hadron collisions or for dijets or vector mesons produced at HERA.        
        
At very low $x$, that is to leading $\log (1/x)$ accuracy, the unintegrated distribution        
becomes independent of the hard scale $\mu$, and so from (\ref{eq:a1}) we have        
\be        
\label{eq:b1}        
f (x, k_t^2, \mu^2) \; \rightarrow \; \left . \frac{\partial}{\partial \ln \lambda^2} \: \left (xg (x,        
\lambda^2) \right ) \right |_{\lambda^2 = k_t^2}.        
\ee        
Clearly (\ref{eq:b1}) cannot remain true as $x$ increases.  Indeed we see that it would give        
negative values for $f$.  Moreover even at low $x$, there are significant subleading        
corrections which, to some level of approximation, modify (\ref{eq:b1}) to the form       
\cite{DDT,KMR}       
\be        
\label{eq:c1}        
f (x, k_t^2, \mu^2) \; \approx \; \left . \frac{\partial}{\partial \ln \lambda^2} \: \left ( xg (x,        
\lambda^2) \: T_g (\lambda, \mu) \right ) \right |_{\lambda^2 = k_t^2},        
\ee        
where $T_g$ is the Sudakov form factor.  In fact (\ref{eq:c1}) is oversimplified.  As        
discussed below, the expression for $f (x, k_t^2, \mu^2)$ is more complicated than  
(\ref{eq:c1}).        
        
The natural framework for unifying the small and large $x$ domains is the CCFM    
formalism based on angular ordering \cite{CCFM}--\cite{JUNG}, which       
follows from colour coherence effects \cite{COH}.  It reduces to the        
leading order DGLAP formalism at moderate $x$ and it embodies the BFKL formalism at       
small $x$.  The unintegrated gluon distribution $f (x, k_t^2, \mu^2)$ satisfies the CCFM       
equation \cite{CCFM,CCFM2} which interlocks the two hard scales $(k_t^2, \mu^2)$ in a  
complicated way.  The       
equation is based on the coherent radiation of gluons, which leads to an angular ordering of       
the gluon emissions along the chain.  The ordering introduces a scale specifying the       
maximum angle of gluon emission, which turns out to be essentially the hard scale $\mu$ of       
the probe.  At moderate $x$ the angular ordering becomes an ordering in the gluon transverse       
momenta and the CCFM equation reduces to DGLAP evolution.  At very small $x$ the       
angular ordering does not provide any constraint on the transverse momenta along the chain       
and, in the leading $\log (1/x)$ approximation, $f (x, k_t^2, \mu^2)$ becomes the $\mu$      
independent distribution which satisfies the       
BFKL equation.  On the other hand, although the dependence on the scale $\mu$ only enters       
at subleading $\log (1/x)$ level, $f$ does depend on $\mu$ through leading $\log \mu^2$       
evolution.        
        
The outline of the paper is as follows.  The angular-ordered  
CCFM equation is introduced in Section 2.  In    
Section 3 we simplify this evolution, yet staying within leading $\log$ accuracy, to        
show that the two-scale distribution $f (x, k_t^2, \mu^2)$ can be obtained in terms of the        
conventional one-scale $g (x, \mu^2)$ distribution.  In this way, we are led to a procedure for    
determining $f (x, k_t^2, \mu^2)$ from a unified DGLAP/BFKL type single-scale evolution    
equation.  This is described in Section 4.  Moreover we are able to extend the formalism to        
incorporate important subleading $\log (1/x)$ effects, which are generated by the so-called        
consistency condition\footnote{Called the kinematic constraint in \cite{KC}.} \cite{KC,AG}        
and which subsume the angular ordering constraint at low $x$. We also extend the 
formalism to    
include the contributions due to the quark distributions.  For comparison, in Section 5 we    
present the pure DGLAP-type approach to determine $f (x, k_t^2, \mu^2)$, in which the    
gluon cascade evolves according to evolution strongly ordered in $k_t$.  Section 6 contains    
sample numerical results for $f (x, k_t^2, \mu^2)$ obtained from the fully unified approach of    
Section 4.  As expected, we have some diffusion of the gluon transverse momenta into the    
region $k_t > \mu$.  This is in contrast to the pure DGLAP-type approximation in which the    
distribution $f (x, k_t^2, \mu^2)$ is limited to the domain $k_t < \mu$.  Finally, Section 7    
contains a summary of the procedure that we have used to determine the unintegrated gluon    
distribution $f (x, k_t^2, \mu^2)$.   
        
\section{The CCFM equation}       
       
The unintegrated gluon distribution satisfies the CCFM evolution equation  
\cite{CCFM,CCFM2}         
based on angular ordering.  In unfolded\footnote{The folded form  
(which actually is the CCFM equation \cite{CCFM,CCFM2}) contains Sudakov and          
non-Sudakov form factors, which arise from the resummation of virtual corrections and         
screen the singularities as $z \rightarrow 1$ and $z \rightarrow 0$ respectively.} form the         
equation is         
\bea         
\label{eq:a2}         
f (x, k_t^2, \mu^2) & = & f_0 (x, k_t^2) \nonumber \\         
& & \nonumber \\         
& + & \frac{\alpha_S}{2 \pi} \, \int_0^1 dz \int \frac{d^2 q}{\pi q^2} \, 
\Theta (1-z-q_0/q)\, 
\biggl [\Theta (z - x) \:     
\Theta (\mu - qz)  \nonumber \\         
& & \nonumber \\         
& & \frac{k_t^2}{k_t^{\prime 2}} \: P (z) f \! \left ( \frac{x}{z}, k_t^{\prime 2}, q^2 \right )     
\: - \: z P(z) \: \Theta (\mu - q)  \: f (x, k_t^2, q^2) \biggr ] \nonumber \\         
& & \nonumber \\         
& - & \; \frac{\alpha_S}{2\pi} \: 2N_C \: \int_x^1 \: \frac{dz}{z} \: \int \: \frac{d^2 q}{\pi         
q^2} \: \Theta (q - q_0) \: \Theta (k_t^2 - q^2) \: f \left ( \frac{x}{z}, k_t^2, q^2 \right ),       
\eea         
where $k_t^\prime \equiv |\mbox{\boldmath $k$}_t + (1 - z) \mbox{\boldmath $q$}|$.  For     
simplicity, $\alpha_S$ is taken outside the integrals, but the scale will be specified       
carefully in the final equations (in particular see (\ref{eq:a15})).   
The driving term $f_0$ is of non-perturbative origin and is       
assumed to contribute only for $k_t^2 < q_0^2$.  The remaining terms contribute in the       
$k_t^2 > q_0^2$ domain.  The driving term thus gives the non-perturbative starting gluon       
distribution         
\be         
\label{eq:aa}         
xg (x, q_0^2) \; = \; \int^{q_0^2} \: \frac{dk_t^2}{k_t^2} \: f_0 (x, k_t^2).        
\ee         
        
This angular-ordered equation for $f$, which embodies both DGLAP and BFKL evolution, is  
shown         
schematically in Fig.~1.  The first term in the square brackets in (\ref{eq:a2}) describes real         
gluon emission with angular ordering imposed.  The term containing $f (x, k_t^2, q^2)$ is         
related to the virtual corrections corresponding to the unfolded Sudakov form factor, whilst         
the last term in (\ref{eq:a2}) represents the virtual corrections which, when resummed, give    
rise to gluon reggeization.  The latter correspond to the BFKL part of the (unfolded) non-   
Sudakov form factor.  Angular ordering along the chain, a portion of which is shown in    
Fig.~2, requires         
\be         
\label{eq:a3}         
z_{n - 1} q_{n - 1} \; < \; q_n \quad\quad {\rm where} \quad\quad q_n \; \equiv \; q_{tn}/(1 -          
z_n)         
\ee         
and $z_n = x_n/x_{n - 1}$.  The angular ordering continues up to a maximum angle with the          
limit expressed in terms of the hard scale via $\Theta (\mu - qz)$.  
%%%%%%%%%%%%%%%%%%%%%%%%%%%%%%%%%%%%%%%%%%%%%%%%%%%%%%%%%%%%%%%%%%%%%%%%%   
\begin{figure}[htb]
\centerline{\epsfig{file=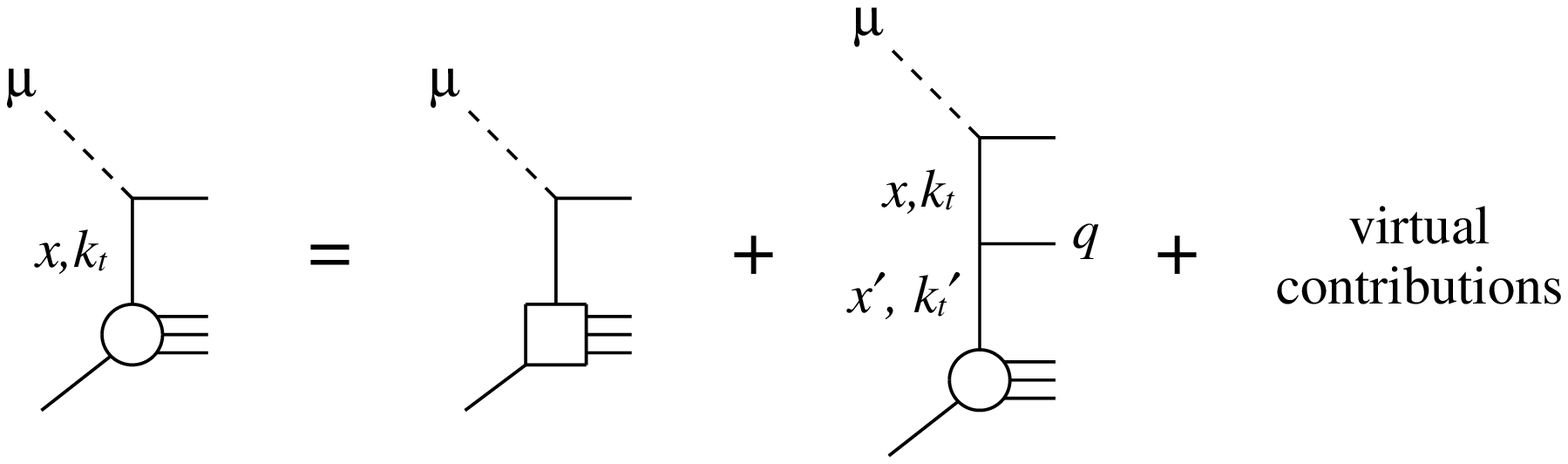,height=6cm,width=14cm}}
\small{Figure 1: Schematic representation of the CCFM evolution equation, (\ref{eq:a2}), for        
the unintegrated gluon distribution $f (x, k_t^2, \mu^2)$.  The variable $q$ is defined by $q        
\equiv q_t/(1 - z)$ where $z = x/x^\prime$, and $\mbox{\boldmath $k$}_t^\prime =        
\mbox{\boldmath $k$}_t + (1 - z) \mbox{\boldmath $q$}$.}
\label{fig:fig1}
\end{figure} 
%%%%%%%%%%%%%%%%%%%%%%%%%%%%%%%%%%%%%%%%%%%%%%%%%%%%%%%%%%%%%%%%%%%%%%%%%    
When $z$ is away from the $z \sim 0$ and $z \sim 1$ domains, angular ordering is equivalent    
to the strong $k_t$ ordering of pure DGLAP LO evolution.  At first sight there appear to 
 be three types of large logarithms in (\ref{eq:a2}).     
First the usual DGLAP logarithms coming from the     
\be    
\label{eq:n1}    
| \mbox{\boldmath $k$}_t^\prime| \; \equiv \; | \mbox{\boldmath $k$}_t + (1 - z)     
\mbox{\boldmath $q$}| \; \ll \; q    
\ee    
domain.  Second there are the BFKL-type $\log (1/x)$ contributions originating from the     
$1/z$ part of the real emission term in (\ref{eq:a2}) and the gluon reggeization contribution.      
These two terms can be combined together in the function     
\bea    
\label{eq:n2}    
F (x, k_t^2) & = & \frac{\alpha_S}{2 \pi} \: 2N_C \: \int \: \frac{d^2 q}{\pi q^2} \: \int_x^1 \:     
\frac{dz}{z} \nonumber \\    
& & \nonumber \\    
& & \left [ \frac{k_t^2}{| \mbox{\boldmath $k$}_t + \mbox{\boldmath $q$}|^2} \: f \left (     
\frac{x}{z}, | \mbox{\boldmath $k$}_t + \mbox{\boldmath $q$}|^2, q^2 \right ) \: - \: \Theta     
(k_t^2 - q^2) \: f \left ( \frac{x}{z}, k_t^2, q^2 \right ) \right ]    
\eea    
where we have assumed $z \ll 1$, in $k_t^\prime$ of (\ref{eq:n1}).  Finally there is a danger     
that in (\ref{eq:a2}) we have a large logarithm from the region $q^2 \ll k_t^2$.  However, we 
see from (\ref{eq:a2}) that the function $f(x,k_t^2,q^2)$ only extends into the region 
$k_t^2>q^2$ as a result of the BFKL small $x$ effects, which are subleading at finite $x$. 
That is, to leading log accuracy, it can be shown, in the so-called ``single-loop''  
approximation, that if the last term in (\ref{eq:a2}) is neglected and $\Theta(\mu-qz)$ is 
replaced by $\Theta(\mu-q)$, then the function $f(x,k_t^2,q^2)$ vanishes for $k_t^2>q^2$. 
Thus we limit the integration regions in (\ref{eq:a2}) to the strongly-ordered domain 
$k_t^{\prime 2}\ll q^2$ for those contributions in which the unintegrated gluon is multiplied 
by the part $\bar{P}(z)$ of the splitting function $P(z)$ which is non-singular at low $z$. 
That is 
\be    
\label{eq:k1}    
\bar{P}(z) = P(z) \: - \: \frac{2N_C}{z}.   
\ee    
It should be noted that the BFKL part, (\ref{eq:n2}), of (\ref{eq:a2}) 
(for which this approximation 
is not justified) is free from singularity as $q\rightarrow 0$, since the potential singularity 
of the real emission term is cancelled by the virtual contribution. 
 %%%%%%%%%%%%%%%%%%%%%%%%%%%%%%%%%%%%%%%%%%%%%%%%%%%%%%%%%%%%%%%%%%%%%%%%%   
\begin{figure}[htb]
\centerline{\epsfig{file=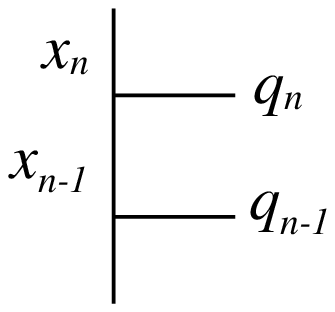,height=4cm,width=3cm}}
\small{Figure 2: A portion of the evolution chain.  Angular ordering requires $z_{n - 1} q_{n -        
1} < q_n$, where $q_n \equiv q_{tn}/(1 - z_n)$ and $z_n = x_n/x_{n - 1}$. }
\label{fig:fig2}
\end{figure} 
%%%%%%%%%%%%%%%%%%%%%%%%%%%%%%%%%%%%%%%%%%%%%%%%%%%%%%%%%%%%%%%%%%%%%%%%%    

So     
finally we have just the large logarithms coming from either $k_t^\prime \ll k_t$ or from $z     
\ll 1$.  Our aim is to develop an approximate treatment of the CCFM equation which     
incorporates both types of large logarithms.    
    
\section{Simplification of the CCFM equation}    
    
To simplify the angular-ordered equation, (\ref{eq:a2}), we rearrange the equation and retain  
only     
terms which generate large logarithms.  To achieve this it is convenient to add and subtract     
the term    
\be    
\label{eq:j1}    
\frac{\alpha_S}{2\pi} \: \int_0^1 \: dz \: \int \: \frac{d^2 q}{\pi q^2} \: \Theta \left ( 1 - z \: - \:     
\frac{q_0}{q} \right ) \: \Theta (\mu - q) \: \Theta (k_t^2 - k_t^{\prime 2}) \: z P(z) \:     
\frac{k_t^2}{k_t^{\prime 2}} \: f (x, k_t^{\prime 2}, q^2)    
\ee    
from the right-hand-side of (\ref{eq:a2}), and to group together contributions containing the     
singular $2 N_C/z$ part of the splitting function    
\be    
\label{eq:j2}    
P(z) \; \equiv \; \bar{P}(z) \: + \: \frac{2N_C}{z}.    
\ee    
In this way we obtain the approximate form    
\bea    
\label{eq:j4}    
\lefteqn{f (x, k_t^2, \mu^2) = f_0 (x, k_t^2) \: + \: \frac{\alpha_S}{2 \pi} \int_0^1 dz \Theta  
(k_t -     
q_0) \Theta \left (\mu - \frac{k_t}{1 - z} \right )}\nonumber \\    
& & \nonumber \\    
& & \times \: \left [\bar{P}(z) \Theta (z - x) \frac{x}{z} g \left (\frac{x}{z}, \left     
(\frac{k_t}{1 - z} \right )^2 \right ) - zP(z) xg \left (x, \left (\frac{k_t}{1 - z} \right )^2 \right )     
\right ] \nonumber \\    
& & \nonumber \\    
&  +& \: \frac{\alpha_S}{2\pi} \int_0^1 dz \int \frac{dq^2}{q^2} \Theta \left ( 1 - z -     
\frac{q_0}{q} \right ) \Theta (\mu - q) \,zP(z) \nonumber \\    
& & \nonumber \\    
&  \times & \left [ \int \frac{dk_t^{\prime 2}}{k_t^{\prime 2}} \Theta (k_t^2 - k_t^{\prime     
2})  f (x, k_t^{\prime 2}, q^2) \frac{q}{2} \delta \left (q - \frac{k_t}{1 - z} \right ) - f (x,     
k_t^2, q^2) \Theta \left (q - \frac{k_t}{1 - z} \right ) \right ] \: + \, F (x, k_t^2), \nonumber \\    
\eea    
where the BFKL-type $\log (1/x)$ contribution $F (x, k_t^2)$ is given by (\ref{eq:n2}).    
The second term on the right-hand-side is the pure DGLAP contribution in the large $x$     
limit.  It comes from the strongly-ordered configuration    
\be    
\label{eq:j5}    
k_t^{\prime 2} \; \equiv \; | \mbox{\boldmath $k$}_t + (1 - z) \mbox{\boldmath $q$}|^2 \; \ll     
\; q^2    
\ee    
in the second term on the right-hand-side of (\ref{eq:a2}).  In this configuration the variable     
$q$ becomes $k_t/(1 - z)$.  We have also made the large $z$ approximation such that    
\be    
\label{eq:j6}    
\bar{P}(z) \: \Theta (\mu - qz) \; \approx \; \bar{P}(z) \: \Theta (\mu - q) \; \approx \;     
\bar{P}(z) \: \Theta \left ( \mu \: - \: \frac{k_t}{1 - z} \right ).    
\ee    
With these approximations we may rewrite the second term using     
\bea    
\label{eq:j7}    
& & \int^{k_t^2/(1 - z)^2} \: \frac{dk_t^{\prime 2}}{k_t^{\prime 2}} \left [ \Theta (z - x) \:     
\bar{P}(z) \: f \left ( \frac{x}{z}, k_t^{\prime 2}, \left ( \frac{k_t}{1 - z} \right )^2 \right ) \: -     
\: zP(z) \: f \left (x, k_t^{\prime 2}, \left (\frac{k_t}{1 - z} \right )^2 \right ) \right ]     
\nonumber \\    
& & \nonumber \\    
& & \quad\quad = \; \Theta (z - x) \: \bar{P}(z) \: \frac{x}{z} \: g \left (\frac{x}{z}, \left     
(\frac{k_t}{1 - z} \right )^2 \right ) \: - \: zP(z) \: xg \left ( x, \left ( \frac{k_t}{1 - z} \right )^2     
\right )    
\eea    
where the upper limit $k_t^2$ of the $dk_t^{\prime 2}$ integration has, to leading $\log     
\mu^2$ accuracy, been replaced by $(k_t/(1 - z))^2$.    
    
Finally, the third term on the right-hand-side of (\ref{eq:j4}) corresponds to the difference    
between (\ref{eq:j1})  and the virtual Sudakov (DGLAP-type) contribution given by the    
second term in the square brackets in (\ref{eq:a2}), that is to the contribution   
\begin{equation}   
\label{eq:k4}   
\frac{\alpha_S}{2 \pi} \: \int_0^1 \: dz \: \int \: \frac{d^2 q}{\pi q^2} \: \Theta \left ( 1 - z    
- \frac{q_0}{q} \right ) \: \Theta (\mu - q) \: zP(z)    
\: \left [ \frac{k_t^2}{k_t^{\prime 2}} \:  
 f (x, k_t^{\prime 2}, q^2) \: - \:  f (x, k_t^2, q^2)    
\right ].   
\end{equation} 
The first integral is evaluated using the strongly-ordered configuration $(k_t^{\prime 2} \ll    
k_t^2, q \approx k_t/(1 - z))$, while the second integration is restricted to the     
region $q (1 - z) > k_t$. 
It should be noted that for $q(1-z)\ll k_t$ the two terms in the integrand of  
(\ref{eq:k4}) cancel, since then $k_t^\prime \approx k_t$. 
 
  The contribution (\ref{eq:k4}) represents the virtual corrections    
which have to be resummed.  The resummation is performed in the Appendix.  We obtain    
\bea    
\label{eq:j8}    
f (x, k_t^2, \mu^2) & = & \frac{\partial}{\partial \ln k_t^2} \: \biggl [T_g (k_t, \mu) \: xg (x,    
k_t^2)  \nonumber \\    
& & \nonumber \\    
& & + \; \int_{k_t^2}^{\mu^2} \: \frac{dq^2}{q^2} \: T_g (q, \mu) \int_{q_0^2}^{k_t^2} \:    
\frac{dk_t^{\prime 2}}{k_t^{\prime 2}} \: \frac{\partial L (x, k_t^{\prime 2}, q^2)}{\partial    
\ln q^2} \biggr ],    
\eea    
where    
\bea    
\label{eq:j10}    
L (x, k_t^2, \mu^2) & = & \frac{\alpha_S}{2 \pi} \: \Theta (k_t - q_0) \: \int_0^1 \: dz \:     
\Theta \left (\mu - \frac{k_t}{1 - z} \right ) \: \left [\bar{P}(z) \: \Theta (z - x) \: \frac{x}{z} \:    
g \left (\frac{x}{z}, \left ( \frac{k_t}{1 - z} \right )^2 \right ) \right . \nonumber \\    
& & \nonumber \\    
& & \quad\quad\quad\quad\quad\quad\quad\quad\quad\quad\quad\quad - \; zP(z) \: xg \left .    
\left (x, \left ( \frac{k_t}{1 - z} \right )^2 \right ) \right ],    
\eea    
and where the Sudakov form factor    
\be    
\label{eq:j9}    
T_g (q, \mu) \; = \; \exp \left ( - \int_{q^2}^{\mu^2} \: \frac{dp^2}{p^2} \:     
\frac{\alpha_S (p^2)}{2 \pi} \: \int_0^{1 - k_t/p} \: dz^\prime \: z^\prime P(z^\prime) \right ).    
\ee    
The cut-off $z^\prime < 1 - k_t/p$ enters on account of the kinematic structure of the real    
emission term, where the upper limit of the $p$ integration is given by $k_t/(1 - z^\prime)$.     
Note that $T_g (q, \mu)$ therefore implicitly depends on $k_t$. 
  The integration limits defining the  Sudakov  
form factor should be understood as arising from the $\Theta$ function constraints    
\be    
\label{eq:j11}    
\Theta (1 - k_t/p) \: \Theta (\mu - p) \: \Theta (p - q).    
\ee    
This implies $T_g (q, \mu) = 1$ if these constraints are not satisfied.  In particular $T_g = 1$    
for $k_t > \mu$ or $q > \mu$.    
   
A nice feature of the result (\ref{eq:j8}) is that the unintegrated gluon $f (x, k_t^2, \mu^2)$ is    
entirely specified in terms of the integrated gluon $xg$.  The next step is to introduce a single    
scale unified equation, which embodies both BFKL and DGLAP type effects, to determine    
$xg$.   
    
\section{Strategy for determining the unintegrated gluon}    
    
We have emphasized that the angular-ordered equation (\ref{eq:j4}) is a \lq two-scale\rq\  
evolution    
equation for $f (x, k_t^2, \mu^2)$.  That is the scales $k_t^2$ and $\mu^2$ are intertwined    
by angular ordering.  In the previous section we have shown how the two-scale unintegrated    
gluon $f (x, k_t^2, \mu^2)$ can be determined once we know the integrated gluon $xg$.     
Here we describe the procedure to obtain $xg$ from a unified evolution equation for a single    
scale auxiliary distribution         
\be         
\label{eq:a13}         
h (x, \mu^2) \; \equiv \; \mu^2 \: \frac{\partial (xg)}{\partial \mu^2} \; = \; \mu^2 \:          
\frac{\partial}{\partial \mu^2} \: \left ( \int^{\mu^2} \frac{dk_t^2}{k_t^2} \: f (x, k_t^2,     
\mu^2) \right ).         
\ee         
If we integrate both sides of equation (\ref{eq:j4}) over $k_t^2$ up to $\mu^2$, differentiate         
with respect to $\log \mu^2$, then we find that $h (x, \mu^2)$ satisfies the evolution equation         
\bea         
\label{eq:a14}         
h (x, \mu^2) & = & \frac{\alpha_S}{2 \pi} \: \int_0^{1 - q_0/\mu} \: dz \: \biggl \{ \Theta         
(z - x) \: \bar{P}(z) \: \frac{x}{z} \: g \left ( \frac{x}{z}, \mu^2 \right ) \biggr . \nonumber \\         
& & \nonumber \\         
& & \biggl . - \; z P(z) \: xg (x, \mu^2) \biggr \} \; + \; F (x, k_t^2 = \mu^2),         
\eea         
where $F (x, k_t^2)$ is defined by (\ref{eq:n2}).  Note     
that the integral in (\ref{eq:a14}) has no singularity close to $z = 1$.     
We can now  derive a relation expressing the     
two-scale unintegrated gluon distribution         
$f (x, k_t^2, \mu^2)$ in terms of the one-scale distribution  
$xg$ (or $h$). From (\ref{eq:j8}) we get the following  
expression for $f (x,k_t^2,\mu)$:     
\bea    
& & f (x,k_t^2,\mu^2) \; = \; T_g(k_t,\mu)h(x,k_t^2) \: + \:    
 T_g(k_t,\mu)\int_0^{1- k_t/\mu} dz \,zP(z)\,\frac{\alpha_S(k_t^2/(1-z)^2)}{2\pi}  
 \, xg(x,k_t^2) \nonumber \\    
& & \nonumber \\    
& & \quad\quad - \;\; {\alpha_S\over 2 \pi}T_g(k_t,\mu) \int_0^{1- q_0/k_t} dz \left[\Theta(z-   
x) \bar P(z){x\over z} g\left({x\over z},k_t^2\right)-P(z)zxg\left(x,k_t^2\right)\right]    
\nonumber \\    
& & \quad\quad + \;\; \Theta(\mu-k_t) \int_0^{1- k_t/\mu}     
dz \: {\alpha_S\over 2 \pi}\:T_g\left({k_t\over 1-z},\mu\right)\left[ \Theta(z-x)\bar    
P(z){x\over z} g\left({x\over z},\left({k_t\over 1-z}\right)^2\right) \right . \nonumber \\    
& & \nonumber \\    
& & \quad\quad\quad\quad\quad\quad\quad\quad\quad\quad\quad\quad\quad\quad\quad\quad    
- \;\; \left . P(z) zxg\left(x,\left({k_t\over 1-z}\right)^2\right)\right] ,    
\label{eq:a15}     
\eea      
where the scale of $\alpha_S$ is taken to be the scale of the appropriate     
gluon, except for the second term on the right-hand-side. 
  We may safely set the $1 - z$ cut-off $q_0/k_t$ to zero, but not the cut-offs     
$k_t/\mu$.     
   
Note that in the leading $\ln (1/x)$ approximation we may set $T_g = 1$ and neglect all the    
integral terms in (\ref{eq:a15}), since they do not generate $\ln (1/x)$ contributions. In this    
approximation the unintegrated gluon is simply   
$$   
f (x, k_t^2, \mu^2) \; = \; h (x, k_t^2),   
$$   
with no dependence on the scale $\mu$.   
   
When obtaining (\ref{eq:a15}) from (\ref{eq:j8}), we have chosen to neglect a contribution    
coming from the derivative of the Sudakov form factor, $T_g (q, \mu)$, with respect to    
$k_t$, which arises from the $k_t$ dependence of the regulator, see (\ref{eq:j9}).  For this    
reason the unintegrated gluon of (\ref{eq:a15}) does not precisely integrate to $xg (x,    
\mu^2)$, although the corrections are subleading in $\log \mu^2$.  The discrepancy is indeed    
negligible at low $x$, but can become of the order of 20\% or so for large values of $x    
\gapproxeq 0.1$.  Rather than complicating (\ref{eq:a15}) by including the derivative, we    
eliminate the discrepancy by changing the regulator in the  form factor (\ref{eq:j9})    
from $k_t/p$ to $q/p$, that is we take the  Sudakov form factor 
\be   
\label{eq:k20}   
T_g (q, \mu) \; = \; \exp \left ( - \int_{q^2}^{\mu^2} \: \frac{dp^2}{p^2} \: \frac{\alpha_S    
(p^2)}{2 \pi} \: \int_0^{1 - q/p} \: dz^\prime \: z^\prime \: P(z^\prime) \right ).   
\ee  
This approximation is justified since in our case either $q=k_t$ or $q\sim k_t$.  
Within this approximation it is evident that the unintegrated gluon (\ref{eq:a15}) integrates    
exactly to $xg (x, \mu^2)$; the sum of the first two terms on the right-hand-side of    
(\ref{eq:a15}) form the total derivative   
\be   
\label{eq:k21}   
k_t^2 \: \frac{\partial}{\partial k_t^2} \: \left [ T_g (k_t, \mu) \: xg (x, k_t^2) \right ] ,   
\ee   
and the integrals of the third and fourth terms cancel each other. 
   
\indent From (\ref{eq:a13}) we see that the integrated gluon distribution $g$ can be     
expressed in terms of $h$, namely         
\be         
\label{eq:a16}         
xg (x, k_t^2) \; = \; xg (x, q_0^2) \: + \: \int_{q_0^2}^{k_t^2} \: \frac{d \mu^2}{\mu^2} \: h         
(x, \mu^2).         
\ee       
Equations (\ref{eq:a14}) and (\ref{eq:a15}), together with (\ref{eq:n2}) and         
(\ref{eq:a16}), form a system of coupled equations.  If we substitute $f$ of (\ref{eq:a15}) into       
the $F$ term on the right-hand-side of (\ref{eq:a14}), and take account of (\ref{eq:a16}), then       
we obtain an integral equation for $h$.  We may solve this equation for the single-scale       
auxiliary distribution $h (x, \mu^2)$, and then compute the two-scale unintegrated gluon $f       
(x, k_t^2, \mu^2)$ from (\ref{eq:a15}).        
        
It is convenient to simplify the integral equation (\ref{eq:a14}) for $h (x, \mu^2)$ using       
approximations which are valid to leading $\log$ accuracy.  To be precise we simplify the       
computation of $F (x, k_t^2)$ of (\ref{eq:n2}).  First instead of allowing the scale $q^2$ of       
$f$ to vary, we note that in (\ref{eq:n2}) the dominant values of $q^2$ are such that $q^2     
\approx k_t^2$.  Moreover we notice that in        
the strongly ordered domain, $k_t^{\prime 2} (\equiv |\mbox{\boldmath $k$}_t +     
\mbox{\boldmath $q$}|^2) \ll k_t^2$, that the first term on the         
right-hand-side of (\ref{eq:n2}) can be simplified using        
\bea        
\label{eq:a18}        
\int \: \frac{d^2 q}{\pi q^2} \: \frac{k_t^2}{k_2^{\prime 2}} \: f \left ( \frac{x}{z},    
|\mbox{\boldmath $k$}_t +  \mbox{\boldmath $q$}|^2, q^2 \right ) & \approx & \int^{k_t^2}    
\: \frac{dk_t^{\prime 2}}{k_t^{\prime 2}} \: f \left ( \frac{x}{z}, k_t^{\prime 2}, k_t^2     
\right ) \nonumber \\        
& & \nonumber \\        
& \equiv & \frac{x}{z} \: g \left ( \frac{x}{z}, k_t^2 \right ),        
\eea        
see (\ref{eq:a1}).  In the remaining contributions to $F (x, k_t^2)$ of (\ref{eq:n2}) we can    
use (\ref{eq:a15}) to approximate $f$ by the first term,        
\be        
\label{eq:a19}        
f (x, k_t^2, q^2) \; \approx \; h (x, k_t^2),    
\ee        
noting that $q^2 \approx k_t^2$ and $T_g (k_t, k_t) = 1$.  The other terms in (\ref{eq:a15})     
give only subleading $\log (1/x)$ contributions to the BFKL        
kernel.  Since this contribution to $F (x, k_t^2)$ goes beyond the strongly-ordered part of the        
kernel, it is also subleading in $\log \mu^2$.  After these approximations, eq.\ (\ref{eq:a14})        
for $h (x, \mu^2)$ for $\mu^2 > q_0^2$ may be written\footnote{In order to be consistent       
with (\ref{eq:a14}), the upper limit of the $z$ integration in the second term on the right-      
hand-side of (\ref{eq:a20}) should be $1 - q_0/q$ rather than 1.  The integrals are, of course,     
regular at $z = 1$ and so the discrepancy is subleading in $\ln \mu^2$.}       
\bea        
\label{eq:a20}        
\lefteqn{h (x, \mu^2) \; = \; h_0 (x, \mu^2) \: + \: \frac{\alpha_S (\mu^2)}{2 \pi} \int_0^1 dz        
\int_{q_0^2}^{\mu^2} \: \frac{dq^2}{q^2} \left \{ \Theta (z - x) \: \bar{P} (z) \: h \left (        
\frac{x}{z}, q^2 \right ) - zP(z) \: h (x, q^2) \right \} }\hspace{8mm}\nonumber \\        
& & \nonumber \\        
&+& \frac{\alpha_S (\mu^2)}{2 \pi} \: 2N_C \int_x^1 \frac{dz}{z}        
\int \frac{dq^2}{q^2} \: \Theta (k_t^{\prime 2} - q_0^2) \: \left \{ \frac{\mu^2}{k_t^{\prime       
2}} \: h \left ( \frac{x}{z}, k_t^{\prime 2} \right ) \: - \: \Theta (\mu^2 -        
q^2) \: h \left ( \frac{x}{z}, \mu^2 \right ) \right \}\quad\quad       
\eea        
where $k_t^{\prime 2} = |\mbox{\boldmath $k$}_t + \mbox{\boldmath $q$}|^2$ with $k_t^2        
= \mu^2$.  The driving term, which arises from the substitution of (\ref{eq:a16}) for $xg$ in        
(\ref{eq:a14}) and (\ref{eq:a18}), is given by        
\be        
\label{eq:a21}        
h_0 (x, \mu^2) \; = \; \frac{\alpha_S (\mu^2)}{2 \pi} \int_0^1 dz \left \{ \Theta (z - x) \: P(z) \:        
\frac{x}{z} \: g \left (\frac{x}{z}, q_0^2 \right ) \: - \: zP(z) \: xg (x, q_0^2) \right \}.        
\ee        
Notice that the strongly-ordered contribution of $F (x, k_t^2 = \mu^2)$ has combined        
with the residual DGLAP contribution in (\ref{eq:a14}) with the effect that $\bar{P}(z)        
\rightarrow P(z)$.        
        
Equation (\ref{eq:a20}) is the single-scale unified BFKL/DGLAP equation for the gluon that        
was proposed in Ref.~\cite{KMS}.  There it was shown that it is straightforward to        
incorporate a major part of the subleading order $\log (1/x)$ (or BFKL) effects by        
imposing a consistency condition to ensure that the virtuality of the exchanged gluon is  
dominated by its transverse momentum squared.  This is achieved by including the theta  
function $\Theta (\mu^2 - zq^2)$ in the real emission contribution in the last term of  
(\ref{eq:a20}).  Other important subleading terms arising from using    
the complete DGLAP splitting function and from the running of $\alpha_S$ are automatically    
included in our framework \cite{KMS}.  This formalism was used to fit to deep inelastic    
scattering data was made and the auxiliary function $h (x, \mu^2)$ was determined    
\cite{KMS}.  It was checked that the corresponding integrated gluon $xg (x, k_t^2)$    
computed from (\ref{eq:a14}) was compatible with the gluons obtained in the MRS, CTEQ    
global parton analyses \cite{PA}.        
       
In \cite{KMS} the contribution of the quark distributions was included in (\ref{eq:a20}) for       
$h (x, \mu^2)$.  To incorporate the quarks in the present analysis we must also include the       
contribution of the singlet quark distribution $\Sigma$ in (\ref{eq:a15}) for $f (x, k_t^2,       
\mu^2)$.  That is we make the replacement       
\be       
\label{eq:a22}       
\bar{P}(z) \: \frac{x}{z} \: g \; \rightarrow \; \bar{P}(z) \: \frac{x}{z} \: g \left (       
\frac{x}{z}, \mu^{\prime 2} \right ) \: + \: P_{gq} (z) \: \frac{x}{z} \: \Sigma \left (       
\frac{x}{z}, \mu^{\prime 2} \right )       
\ee       
in the real emission part of the third and fourth  
terms on the right-hand-side of (\ref{eq:a15}), where $\mu^\prime$       
is the appropriate scale.  Recall that $P (z) \equiv P_{gg} (z)$.  In the second 
 term, and in the virtual part of the third and fourth terms, we make the     
replacement       
\be       
\label{eq:a23}       
zP (z) \: xg \; \rightarrow \; \left (zP(z) \: + \: 2n_f \: zP_{qg}(z) \right ) \:  
xg (x, \mu^{\prime 2}),    
\ee       
where $n_f$ is the number of active flavours, and the scale 
$\mu^{\prime 2}=k_t^2$ or $k_t^2/(1-z)^2$ as appropriate.  Finally  
we have to modify the Sudakov form     
factor so that (\ref{eq:k20}) becomes       
\be    
\label{eq:a26}    
T_g (q, \mu) \; = \; \exp \left (- \int_{q^2}^{\mu^2} \: \frac{\alpha_S (p^2)}{2 \pi} \:     
\frac{dp^2}{p^2} \int_0^{1 - q/p} z^\prime \left [P(z^\prime) \: + \: \sum_q \: P_{qg}    
(z^\prime) \right ] \: dz^\prime \right ) .   
\ee    
       
The above procedure allows the determination of the approximate solution $f (x, k_t^2,       
\mu^2)$ of the CCFM equation, which incorporates both a full (or so-called        
\lq\lq all-loop\rq\rq\ \cite{BRW,MBRW}) resummation of leading $\ln (1/x)$ contributions,       
as well as the resummation of leading $\ln \mu^2$ contribution, and the inclusion of       
dominant subleading $\ln (1/x)$ terms.  \\       
       
\section{The pure DGLAP limit}       
       
It is informative to compare the predictions above for the unintegrated gluon $f (x, k_t^2,       
\mu^2)$ with those obtained in the DGLAP (or so-called \lq\lq single-loop\rq\rq\       
\cite{BRW,MBRW}) approximation, in which $\Theta (\mu - qz)$ in (\ref{eq:a2}) is       
replaced by $\Theta (\mu - q)$, and the last term in (\ref{eq:a2}) is neglected.  After making    
these modifications we repeat the procedures of Sections 3 and 4 and obtain the DGLAP form   
\bea   
\label{eq:a27}   
\lefteqn{f (x, k_t, \mu^2) \; = \; \int_x^{1 - k_t/\mu} \: dz \:  
\frac{\alpha_S(k_t^2/(1-z)^2)}{2 \pi} \: T_g \left (    
\frac{k_t}{1 - z}, \mu \right ) \: P(z) \: \frac{x}{z} \: g \left ( \frac{x}{z}, \left (    
\frac{k_t}{1 - z} \right )^2 \right ) }\nonumber \\   
& & \nonumber \\   
& + & \!\!\!\!\!\int_0^{1 - k_t/\mu}\!\!dz \, zP(z) \frac{\alpha_S(k_t^2/(1-z)^2)}{2 \pi} 
\left [T_g (k_t, \mu) \,    
 xg (x, k_t^2)  -  T_g \left ( \frac{k_t}{1 - z}, \mu \right ) \,    
 xg \left (x, \left ( \frac{k_t}{1 - z} \right )^2 \right ) \right ],    
\nonumber \\   
\eea   
with $f = 0$ if $k_t > \mu$.  Apart from the last term, this is the equation for the unintegrated    
gluon introduced in Ref.~\cite{KMR}.  The last term, which is only non-zero on account of    
different scales, introduces subleading corrections.  Its inclusion improves the accuracy of    
the integration of $f (x, k_t^2, \mu^2)$ to reproduce $xg (x, \mu^2)$, see (\ref{eq:a1}).  Note    
that the DGLAP or \lq single-loop\rq\  unintegrated gluon vanishes for $k_t \geq \mu$, as    
indeed it must \cite{BRW,MBRW}. In the results presented below we include the quark    
contributions as described in (\ref{eq:a22})--(\ref{eq:a26}).   
   
It is informative to see how the full equation, (\ref{eq:a15}), for $f (x, k_t^2, \mu)$ reduces   
to    
the DGLAP limit (\ref{eq:a27}).  A crucial observation is that in the DGLAP domain $(k_t <    
\mu)$ it is possible, within the leading $\ln (1/x)$ and $\ln (\mu^2)$ accuracy, to replace    
$\bar{P}(z)$ by $P(z)$ in (\ref{eq:a15}).  Thus, {\it provided} $k_t < \mu$, we obtain the    
following more symmetric formula   
\bea    
& & f(x,k_t^2,\mu^2) \;= \; T_g(k_t,\mu)h(x,k_t^2) \: + \:     
T_g(k_t,\mu)\int_0^{1- k_t/\mu} dz \,zP(z)\,\frac{\alpha_S(k_t^2/(1-z)^2)}{2 \pi} \:  
xg(x,k_t^2) \nonumber \\    
& & \nonumber \\    
& & \quad\quad\quad - \;\; \frac{\alpha_S(k_t^2)}{2\pi}\, T_g(k_t,\mu) \int_0^{1- q_0/k_t}  
dzP(z)    
\left[ \Theta(z-x){x\over z} g\left({x\over z},k_t^2\right)-zxg\left(x,k_t^2\right)\right]    
\nonumber \\    
& & \nonumber \\    
& & \quad\quad\quad + \;\; \Theta(\mu-k_t) \int_0^{1- k_t/\mu}     
dz \: {\alpha_S\over 2 \pi} \: T_g\left({k_t\over 1-z},\mu\right)P(z)\left[ \Theta(z-x){x\over    
z} g\left({x\over z},\left({k_t\over 1-z}\right)^2\right) \right . \nonumber \\    
& &  \quad\quad\quad\quad\quad\quad\quad\quad\quad\quad\quad\quad\quad\quad\quad\quad    
- \;\; \left . zxg\left(x,\left({k_t\over 1-z}\right)^2\right)\right] .    
\label{eq:a115}     
\eea    
In the DGLAP limit the first and third terms on the right-hand-side of (\ref{eq:a115}) exactly    
cancel\footnote{Strictly speaking the cancellation is only exact when $q_0 \rightarrow 0$ in    
(\ref{eq:a115}).}; they simply represent the DGLAP equation for $h (x, k_t^2)$ of    
(\ref{eq:a13}).  Thus (\ref{eq:a115}) reduces to (\ref{eq:a27}).   
   
\section{Numerical evaluation of the unintegrated gluon}   
   
In Fig.~3 we show the $k_t$ distributions of three different unintegrated gluons at 
each of four different values of $x$ at a hard scale $\mu^2=100\ \textrm{GeV}^2$. 
Briefly 
\begin{list}{(\roman{bean})}{\usecounter{bean}\setlength{\rightmargin}{\leftmargin}} 
\item the continuous curves are the gluons $f(x,k_t^2,\mu^2)$ obtained from 
(\ref{eq:a15}), with the quark terms included, using the auxiliary function $h(x,k_t^2)$ 
of Ref.~\cite{KMS}, which was itself obtained from a fit to deep inelastic scattering 
data using a unified BFKL/DGLAP equation.  (The curves have been smoothed in the  
transition region $k_t\sim \mu$.) 
\item the dot-dashed curves show $h(x,k_t^2)$ itself \cite{KMS}, which is independent 
of $\mu^2$, 
\item the dashed curves show $f(x,k_t^2,\mu^2)$ calculated from the pure DGLAP equation, 
(\ref{eq:a115}), using in this case the auxiliary function $h(x,k_t^2)$ obtained in 
\cite{KMS} from pure DGLAP evolution from exactly the same starting distributions 
($xg(x,q_0^2)$ etc) as those found in the unified fit. 
\end{list} 
%%%%%%%%%%%%%%%%%%%%%%%%%%%%%%%%%%%%%%%%%%%%%%%%%%%%%%%%%%%%%%%%%%%%%%%%%   
\begin{figure}[htb]
\centerline{\epsfig{file=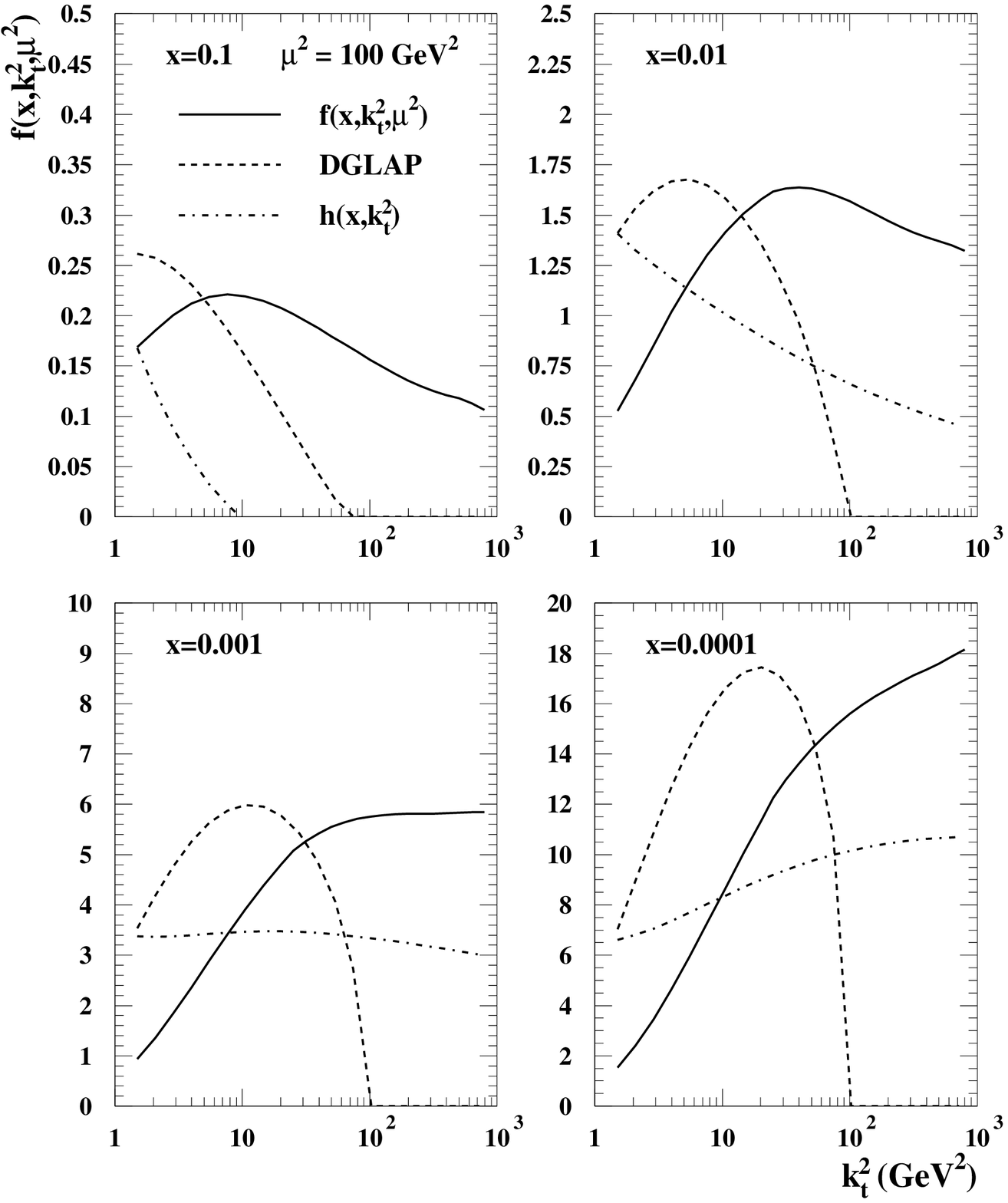,height=15cm,width=13cm}}
\small{Figure 3: The continuous curves show the $k_t$ dependence of the unintegrated 
gluon distribution $f(x,k_t^2,\mu^2)$ for $\mu^2=100\ \textrm{GeV}^2$.  For comparison 
we also show the input auxiliary function $h(x,k_t^2)$ (dot-dashed curves) \cite{KMS} 
and the $k_t$ dependence coming from pure DGLAP evolution (dashed curves).}
\label{fig:fig3}
\end{figure} 
%%%%%%%%%%%%%%%%%%%%%%%%%%%%%%%%%%%%%%%%%%%%%%%%%%%%%%%%%%%%%%%%%%%%%%%%%    
 
Note that the third set of gluons are shown solely to illustrate the difference between 
two types of evolution.  The gluons of the third set have not been constrained by a  
fit to the data, so should not be regarded as realistic.  
   
In the pure DGLAP case, (iii), we see that the distributions are confined to the domain 
$k_t<\mu$, as anticipated from strong ordering.  On the other hand the distributions 
$f(x,k_t^2,\mu^2)$ obtained in the unified BFKL/DGLAP framework develop a more and  
more 
extensive $k_t>\mu$ tail as $x$ decreases.  At small $k_t$ and low $x$ the magnitude 
of the unintegrated gluon calculated from the unified scheme is about a factor of two 
less than that of the gluon coming from the pure DGLAP approach of case (iii). This 
is due to the imposition of the consistency constraint in case (i) which suppresses 
the magnitude of the gluon.  If this constraint were absent the distributions of 
cases (i) and (iii) would not be that different. 
We note that the auxiliary function $h$ of case (ii) remains different from 
the unintegrated gluon $f$ of case (i) down to 
very small values of $x$.

For $k_t > \mu$ we see that $f$ is greater than $h$, whereas the DGLAP-driven unintegrated  
gluon vanishes, as it must.  In this domain inspection of (\ref{eq:a14}) and (\ref{eq:a15}) 
shows that $f$ comes purely from the BFKL contribution,  
\be 
\label{eq:36} 
f (x, k_t^2, \mu^2) \; = \; F (x, k_t^2). 
\ee 
On the other hand $h$ is smaller than $f$ due to the negative contribution of the integral term  
in (\ref{eq:a14}).  The latter is a DGLAP contribution which is ruled out when angular  
ordering is imposed.  It is negative because the $2N_C/z$ contribution has been subtracted  
from the real emission contribution, but not from the virtual term.  We see that for $x < 0.01$  
that there is about a factor of 2 discrepancy between $h$ and the true unintegrated gluon $f$. 
 
\section{Summary}   
   
Here we have addressed the issue of obtaining a reliable determination of the (scale    
dependent) gluon distribution $f (x, k_t^2, \mu^2)$, unintegrated over the gluon transverse    
momentum $k_t$, where $\mu$ denotes the hard scale of the probe.  In the leading $\log    
(1/x)$ approximation the distribution is given simply by the derivative of the unintegrated    
gluon with respect to its scale $\lambda = k_t$, see (\ref{eq:b1}), and satisfies the BFKL    
equation.  We correct this simple relation by going beyond the leading $\log (1/x)$    
approximation to include both subleading contributions and DGLAP effects.  The final result    
for $f (x, k_t^2, \mu^2)$ is given in (\ref{eq:a15}).  To obtain this result we use the    
appropriate gluon cascade formalism based on angular ordering, which leads to the CCFM    
equation embodying both BFKL and DGLAP evolution.  It is important to note that the    
CCFM equation gives a well-defined framework to calculate the very quantity that we seek:     
the unintegrated gluon distribution $f (x, k_t^2, \mu^2)$.  Using this formalism we devise a    
procedure to determine $f (x, k_t^2, \mu^2)$ from the integrated gluon distribution $xg (x,    
q^2)$, its derivative $h (x, q^2)$ and the Sudakov form factor $T_g (q, \mu)$, cf.\    
(\ref{eq:a15}).  An important ingredient is the solution of a (single scale) evolution equation    
for $h (x, q^2)$ which embodies both BFKL and DGLAP effects.  From the low $x$    
viewpoint it includes subleading effects from (i) the consistency constraint which limits the    
available phase space to the region in which the virtuality of the exchanged gluon is    
dominated by its transverse momentum squared, (ii) DGLAP effects generated by that part of    
the splitting function $P_{gg}(z)$ which is not singular in the limit $z \rightarrow 0$, (iii) the    
inclusion of the quark contribution and, (iv) allowing the coupling $\alpha_S$ to run and    
depend on the local scale(s) characteristic of the vertices of the cascade.   
   
We presented sample results to show that the structure of the $k_t$ distribution of the gluon    
$f (x, k_t^2, \mu^2)$ can be significantly different from that of $h (x, k_t^2)$, down to very    
small values of $x$.  There are important consequences for the description of hadron-initiated    
hard processes in which the $k_t$ of the gluon is probed locally.   
   
\vspace{9mm}   
\noindent{\Large\textbf{Appendix}}    
\vspace{4mm} 
   
\noindent Eq.~(\ref{eq:j8}) for the unintegrated gluon $f (x, k_t^2, \mu^2)$ is obtained from    
(\ref{eq:j4}) by resumming the virtual corrections given in the third term on the    
right-hand-side of (\ref{eq:j4}).  Here we show how the resummation is performed.  First we    
note that this virtual correction term can be written as a derivative, that is   
\bea   
\label{eq:append1}   
&&\!\!\!\!\!\!\!\!\! 
\frac{\alpha_S}{2 \pi} \int_0^1 dz \: zP(z) \left [\int \frac{dk_t^{\prime 2}}{k_t^{\prime    
2}} \Theta (k_t^2 - k_t^{\prime 2}) f (x, k_t^{\prime 2}, q^2) \frac{q}{2} \delta \left (q -    
\frac{k_t}{1 - z} \right ) - f (x, k_t^2, q^2) \Theta \left ( q - \frac{k_t}{1 - z} \right ) \right ]    
\nonumber \\   
& & \nonumber \\   
& & \quad\quad\quad\quad\quad\quad\quad\quad\quad\quad\quad = \; - k_t^2 \:    
\frac{\partial}{\partial k_t^2} \: \left [A (k_t, q) \: R (x, k_t^2, q^2) \right ]   
\eea   
where   
\bea   
\label{eq:append2}   
R (x, k_t^2, q^2) & = & \int^{k_t^2} \: \frac{dk_t^{\prime 2}}{k_t^{\prime 2}} \: f (x,    
k_t^{\prime 2}, q^2), \\   
& & \nonumber \\   
\label{eq:append3}   
A (k_t, q) & = & \frac{\alpha_S}{2 \pi} \: \int_0^1 \: dz \: zP(z) \: \Theta (1 - z - k_t/q).   
\eea   
The function $R(x,k_t^2,q^2)$ has a simple physical meaning.  It is the gluon 
distribution for fixed impact parameter $b\sim 1/k_t$ at scale $q$.  Note that if $b=1/q$ then  
the 
distribution $R$ reduces to the integrated gluon $xg\left(x,q^2\right)$.  Using  
(\ref{eq:append1}) 
we see that (\ref{eq:j4}) can be expressed as an integro-differential equation for $R$   
\bea   
\label{eq:append4}   
& & f (x, k_t^2, \mu^2) \; \equiv \; k_t^2 \: \frac{\partial R (x, k_t^2, \mu^2)}{\partial k_t^2}    
\nonumber \\   
& & \nonumber \\   
& & \; = \; L (x, k_t^2, \mu^2) \: + \: F (x, k_t^2) \: - \: k_t^2 \: \frac{\partial}{\partial k_t^2}    
\: \int_{k_t^2}^{\mu^2} \: \frac{dq^2}{q^2} \: A (k_t, q) \: R (x, k_t^2, q^2) ,   
\eea   
where $L$ and $F$ are defined by (\ref{eq:j10}) and (\ref{eq:n2}) respectively.   
   
In order to solve (\ref{eq:append4}) for $R$ we integrate both sides over $dk_t^2/k_t^2$ up    
to $k_t^2$ and obtain the following integral equation   
\bea   
\label{eq:append5}   
R (x, k_t^2, \mu^2) \; = \; xg_0 (x) & + & \int_{q_0^2}^{k_t^2} \: \frac{dk_t^{\prime    
2}}{k_t^{\prime 2}} \: \left [ L (x, k_t^{\prime 2}, \mu^2) \: + \: F (x, k_t^{\prime 2}) \right    
] \nonumber \\   
& & \nonumber \\   
& - & \int_{k_t^2}^{\mu^2} \: \frac{dq^2}{q^2} \: A (k_t, q) \: R (x, k_t^2, q^2).   
\eea   
\noindent From (\ref{eq:a13}) and (\ref{eq:a14}) we see, if $\mu = k_t$, that the first two  
terms on the    
right-hand-side of (\ref{eq:append5}) are just $xg (x, k_t^2)$.  The solution of    
(\ref{eq:append5}) may be therefore written   
\bea   
\label{eq:append6}   
R (x, k_t^2, \mu^2) & = & T_g (k_t, \mu) \: xg (x, k_t^2) \nonumber \\   
& & \nonumber \\   
& & + \; \int_{k_t^2}^{\mu^2} \: \frac{dq^2}{q^2} \: T_g (q, \mu) \: q^2 \:    
\frac{\partial}{\partial q^2} \: \left ( \int_{q_0^2}^{k_t^2} \: \frac{dk_t^{\prime    
2}}{k_t^{\prime 2}} \: L (x, k_t^{\prime 2}, q^2) \right ) ,   
\eea   
where the Sudakov form factor   
\be   
\label{eq:append7}   
T_g (q, \mu) \; = \; \exp \: \left ( - \int_{q^2}^{\mu^2} \: \frac{dp^2}{p^2} \: A (k_t, p) \right    
)   
\ee   
is in agreement with (\ref{eq:j9}).  Eq.~(\ref{eq:a15}) then follows from (\ref{eq:append6})    
after differentiation by $\partial/\partial \ln k_t^2$.   
     
\vspace{9mm}   
\noindent{\Large\textbf{Acknowledgements}}    
\vspace{4mm} 
   
\noindent We thank Misha Ryskin for continual advice and encouragement 
throughout the course of this work.  JK and AMS also thank Grey College 
and the Physics Department of the University of Durham for their warm 
hospitality.  This research has been partially supported by PPARC, by 
a joint KBN-British Council award and by 
the EU Framework TMR programme, 
contract FMRX-CT98-0194.

%\newpage        
%\noindent {\large \bf Figure Captions}        
%\begin{itemize}        
%\item[Fig.\ 1.] Schematic representation of the CCFM evolution equation, (\ref{eq:a2}), for        
%the unintegrated gluon distribution $f (x, k_t^2, \mu^2)$.  The variable $q$ is defined by $q        
%\equiv q_t/(1 - z)$ where $z = x/x^\prime$, and $\mbox{\boldmath $k$}_t^\prime =        
%\mbox{\boldmath $k$}_t + (1 - z) \mbox{\boldmath $q$}$.        
%        
%\item[Fig.\ 2.] A portion of the evolution chain.  Angular ordering requires $z_{n - 1} q_{n -        
%1} < q_n$, where $q_n \equiv q_{tn}/(1 - z_n)$ and $z_n = x_n/x_{n - 1}$. 
% 
%\item[Fig.\ 3.] The continuous curves show the $k_t$ dependence of the unintegrated 
%gluon distribution $f(x,k_t^2,\mu^2)$ for $\mu^2=100\ \textrm{GeV}^2$.  For comparison 
%we also show the input auxiliary function $h(x,k_t^2)$ (dot-dashed curves) \cite{KMS} 
%and the $k_t$ dependence coming from pure DGLAP evolution (dashed curves). 
%\end{itemize}        

\end{document}